\documentclass[journal,article,submit,pdftex,moreauthors]{mdpi} 
\firstpage{1} 
\pubvolume{1}
\issuenum{1}
\articlenumber{0}
\pubyear{2022}
\copyrightyear{2022}
\datereceived{9 March 2022} 
\dateaccepted{11 April 2022} 
\datepublished{} 
\hreflink{https://doi.org/} 

\Title{Observational constraints and some toy models in $f(Q)$ gravity with bulk viscous fluid}

\TitleCitation{Observational constraints and some toy models in $f(Q)$ gravity with bulk viscous fluid}

\Author{Sanjay Mandal $^{1,\dagger}$\orcidA{}, Abhishek Parida $^{2,\ddagger}$ and P.K. Sahoo $^{3,*}$\orcidC{}}

\AuthorNames{Sanjay Mandal, Abhishek Parida and P.K. Sahoo}

\AuthorCitation{Mandal, S.; Parida, A.; Sahoo, P.K.}

\address{%
$^{1}$ \quad Department of Mathematics, Birla Institute of Technology and Science-Pilani, Hyderabad Campus, Hyderabad-500078, India; sanjaymandal960@gmail.com\\
$^{2}$ \quad International College of Liberal Arts, Yamanashi Gakuin University, Yamanashi 400-0805 Japan; abhishekparida22@gmail.com\\
$^{3}$ \quad Department of Mathematics, Birla Institute of Technology and Science-Pilani, Hyderabad Campus, Hyderabad-500078, India; pksahoo@hyderabad.bits-pilani.ac.in}

\corres{Correspondence: pksahoo@hyderabad.bits-pilani.ac.in}

\preto{\abstractkeywords}{\nolinenumbers}
\abstract{The standard formulation of general relativity fails to describe some recent interests in the universe. It impels us to go beyond the standard formulation of gravity.  The $f(Q)$ gravity theory is an interesting modified theory of gravity, where the gravitational interaction is driven by the nonmetricity $Q$. This study aims to examine the cosmological models with the presence of bulk viscosity effect in the cosmological fluid within the framework of $f(Q)$ gravity. We construct three bulk viscous fluid models, i.e. (i) for the first model, we assuming the Lagrangian $f(Q)$ as linear dependence on $Q$, (ii) for the second model the Lagrangian $f(Q)$ as a polynomial functional form, and (iii) the Lagrangian $f(Q)$ as a logarithmic dependence on $Q$. Furthermore, we use 57 points of Hubble data and 1048 Pantheon dataset to constraint the model parameters. Then, we discuss all the energy conditions for each model, which helps us to test the self-consistency of our models. Finally, we present the profiles of the equation of state parameters to test the models' present status.}


\begin{document}

\section{Introduction}\label{sec1}

The current accelerated expansion scenario of the universe is still far away from our fundamental understanding \cite{Riess/1998, 1a,1b,1c}. This problem motivated the research community to go beyond Einstein's general theory of relativity (GR) to describe the candidates responsible for the present scenarios, namely dark energy and dark matter. In GR, the addition of cosmological constant in field equations helps us to understand the unknown form of energy, but it faces some issues such as coincidence problem, the fine-tuning problem, its effects are only observed at cosmological scales (the current accelerated expansion phase) instead of Planck scales \cite{Carlip/2019}. Therefore, in the last few decades, several alternative proposals have been presented in the literature to overcome the current issues of the universe and explore the new insights in the universe.

Recently, a novel proposal has been proposed by Jimenez et al. \cite{Jimenez/2018}, namely symmetric teleparallel gravity or $f(Q)$ gravity, where the fundamental of gravitational interaction is described by the nonmetricity $Q$. Studies on $f(Q)$ gravity are developed by a large number and the observational constraints to face it against standard GR formalism.

An interesting work on symmetric teleparallel gravity was done by Lazkoz et al. \cite{Lazkoz/2019}, where a set of $f(Q)$ function were constraint. To do that, they reformulated the Lagrangian $f(Q)$ as a function of redshift $z$ and discussed the observational constraints using data from gamma-ray bursts, early-type galaxies, cosmic microwave background, type Ia supernovae, baryon acoustic oscillations, and quasars. A relevant study on $f(Q)$ gravity was done by Mandal et al. \cite{Mandal/2020} to understand its behavior using energy conditions, where energy conditions for $f(Q)$ gravity were derived and tested the self-stability of two $f(Q)$ models. Also, they used the parametrization technique and cosmographic idea to constraints three cosmographic sets of functions using the largest Pantheon supernovae data through statistical analysis in $f(Q)$ gravity \cite{Mandal/2020a}. Indeed, some of the interesting studies were done in this modified theory (see details in \cite{Harko/2018, 7a, 7b,7c,7d}).

In literature, most of the cosmological models are considered perfect fluid as the matter content of the universe and discussed its evolution. It is equally important to investigate more physically reliable models such as the dissipative phenomena in the form of bulk viscous might affect the evolution history of the universe, which has been discussed regularly. However, the presence of bulk viscous in a homogeneous and isotropic universe is capable of modifying the background dynamics. The viscous cosmological scenarios come into the picture after the introduction of relativistic thermodynamics. The standard expression for relativistic viscosity was obtained by Eckart in 1940 \cite{bv1}. Later, its cosmological applications are investigated by Weinberg \cite{bv2} and Treciokas and Ellis \cite{bv3}. Then, the bulk viscosity is used to explore the universe's early evolution, whether it is in the case of the neutrino decoupling process or the inflation epoch. In the 1970s, various cosmological applications of bulk viscous imperfect fluid have been discussed \cite{bv4,bv5,bv7,bv8}. An imperfect fluid with bulk viscosity can also describe the acceleration of the universe without the presence of a scalar field or cosmological constant \cite{wc1,wc2,wc3,wc4}. Further, an inflation process can be explained with viscous pressure has been proposed in 1980s \cite{bv9,bv10,bv11,bv12}. Also, it is well-know that, the Israel-Stewart approach is used in many studies to explore the causality with the bulk viscous matter [for instance one can see \cite{IS1, IS2, IS3}]. Moreover, the bulk viscous in the cosmic fluid produces effective pressure when it expands faster than the system to restore its thermal stability \cite{bv13}. The effective pressure can be treated as bulk viscosity. In an accelerated expanding universe it may be natural to assume the possibility that the expansion process is actually a collection of states out of thermal equilibrium in a small fraction of time due to the existence of a bulk viscosity \cite{bv14}.  In addition, the bulk viscosity produces effective negative pressure, which could be considered a suitable candidate for the current accelerated expansion of the universe \cite{wc4,bv18,bv19,bv20,bv21}. It is worthy of mentioning here that bulk viscous can describe both dark matter and dark energy simultaneously \cite{wc4,bv15,bv17}. The possibility of violating the dominant energy condition (DEC) is a well-known result of the FRW cosmological solutions, which correspond to a universe filled with perfect fluid and bulk viscous stresses. The debate over the bulk viscosity is reasonable and practical when it comes to the late time expansion of the universe as we do not know the nature of dark energy and dark matter. Such a possibility only has been investigated in the context of the primordial universe and non-singular model searches. Therefore, we aim to study the current accelerated expansion of the universe by introducing the bulk viscous in the cosmic fluid under the framework of the modified theory of gravity.

In this study, we have added the bulk viscous effect in the cosmological fluid to explore the universe's present scenario. As it is well known that the late-time cosmic acceleration is driven by the negative pressure, and the bulk viscous fluid also negatively affects the pressure. This might help us understand the universe's current expansion through $f(Q)$ gravity. The advantage of working on this theory over the teleparallel theory is that the $f(Q)$ connections admit freely specifiable functions, and these free functions are promoted to true connection degrees of freedom, whereas teleparallel connections are completely fixed. They not only enter the metric field equations but also possess their equations of motion, and they genuinely influence the gravitational field. Therefore, it is adequate to say that $f(Q)$ cosmology is more valuable than $f(T)$ cosmology at the background level because of these additional degrees of freedom. The above case is true for flat and non-flat spatial cases. A possible explanation is that while both connections in $f(T)$ and $f(Q)$ gravity respect homogeneity and isotropy and are flat, the $f(T)$ connection has also to satisfy the conditions $Q_{\alpha\mu\nu} = 0$. These are $40$ independent equations, as opposed to the $24$ independent equations the $f(Q)$ connection has to satisfy because of the conditions $T^{\alpha}_{\mu\nu}=0$. Hence, the $f(T)$ connection is generally more restricted than the $f(Q)$ connection. In literature, many studies have been done on the bulk viscous fluid to deal with the present issues of the universe. For instance, Arora et al. \cite{Arora/2020} studied bulk viscosity models in $f(R, T)$ gravity, where stability analysis of the cosmological models have been examined by focusing on the current phase as well as tested against the observational data from Union 2.1 type Ia supernovae and Hubble data. The cosmic expansion is studied with matter creation and bulk viscosity in \cite{Cardenas/2020}. Here, we are focusing on investigating the stability analysis of the bulk viscous fluid models in modified $f(Q)$ gravity in correspondence with the universe's present scenario.

This article is organized as follows: in Section \ref{sec2}, we briefly discuss the $f(Q)$ gravity framework and derive the field equations for the bulk viscous fluid. Energy conditions are the greatest tool to test the self-stability of the cosmological models. We discuss the energy conditions for $f(Q)$ gravity in Section \ref{sec3}. In Section \ref{sec4}, we use Markov Chain Monte Carlo (MCMC) analysis to estimate the coefficients in the expression of $H(z)$. There the 57 Hubble data points and  1048 Pantheon supernovae dataset are used for simulation. The energy conditions of three viscous fluid models are examined in Section \ref{sec5}. Also, the equation of state parameters discussed for each model in Section \ref{sec5}. Finally, we discuss the final outcomes and future perspectives in Section \ref{sec6}.

\section{Geometrical Overview}\label{sec2}

Here, we have considered the action for symmetric teleparallel gravity is given by \cite{Jimenez/2018}
\begin{equation}
\label{1}
\mathcal{S}=\int\frac{1}{2}\,f(Q)\sqrt{-g}\,d^4x+\int \mathcal{L}_m\,\sqrt{-g}\,d^4x\,
\end{equation}
where $f(Q)$ represents the function form of Q, $g$ is the determinant of the metric $g_{\mu\nu}$, and $\mathcal{L}_m$ is the matter Lagrangian density.\\
The non-metricity tensor and its traces can be written as\\
\begin{equation}
\label{2}
Q_{\lambda\mu\nu}=\bigtriangledown_{\lambda} g_{\mu\nu}
\end{equation}
\begin{equation}
\label{3}
Q_{\alpha}=Q_{\alpha}\;^{\mu}\;_{\mu},\; \tilde{Q}_\alpha=Q^\mu\;_{\alpha\mu}
\end{equation}
Also,the non-metricity tensor helps us to write the  superpotential as 
\begin{equation}
\label{4}
P^\alpha\;_{\mu\nu}=\frac{1}{4}\left[-Q^\alpha\;_{\mu\nu}+2Q_{(\mu}\;^\alpha\;_{\nu)}+Q^\alpha g_{\mu\nu}-\tilde{Q}^\alpha g_{\mu\nu}-\delta^\alpha_{(\mu}Q_{\nu)}\right]
\end{equation}
where the trace of non-metricity tensor \cite{Jimenez/2018} has the form
\begin{equation}
\label{5}
Q=-Q_{\alpha\mu\nu}\,P^{\alpha\mu\nu}
\end{equation}
Again, by definition, the energy-momentum tensor for the fluid description of the spacetime cab be written as
\begin{equation}
\label{6}
T_{\mu\nu}=-\frac{2}{\sqrt{-g}}\frac{\delta\left(\sqrt{-g}\,\mathcal{L}_m\right)}{\delta g^{\mu\nu}}
\end{equation}
Now, one can write the motion equations by varying the action \eqref{1} with respect to metric tensor $g_{\mu\nu}$, which can be written as
\begin{equation}
\label{7}
\frac{2}{\sqrt{-g}}\bigtriangledown_\gamma\left(\sqrt{-g}\,f_Q\,P^\gamma\;_{\mu\nu}\right)+\frac{1}{2}g_{\mu\nu}f
+f_Q\left(P_{\mu\gamma i}\,Q_\nu\;^{\gamma i}-2\,Q_{\gamma i \mu}\,P^{\gamma i}\;_\nu\right)=-T_{\mu\nu},
\end{equation}

where $f_Q=\frac{df}{dQ}$. Also varying \eqref{1} with respect to the connection,one obtains
\begin{equation}
\label{8}
\bigtriangledown_\mu \bigtriangledown_\nu \left(\sqrt{-g}\,f_Q\,P^\gamma\;_{\mu\nu}\right)=0.
\end{equation}
The FLRW line element is given by
\begin{equation}
\label{9}
ds^2=-dt^2+a^2(t)\delta_{\mu\nu}dx^{\mu}dx^{\nu},
\end{equation}
where $a(t)$ is the scale factor of the universe. For this line element the trace of non-metricity tensor takes the form as follow
\begin{align*}
Q=6H^2.
\end{align*} 
We take the energy-momentum tensor of the cosmological fluid which is given by
\begin{equation}
\label{10}
T_{\mu\nu}=(\bar{p}+\rho)u_{\mu}u_{\nu}+\bar{p}g_{\mu\nu},
\end{equation}
where $\bar{p}=p-3\lambda H$ and $\lambda>0$ represents the pressure with viscous fluid and $\rho$ represents the energy density. The physical unit of cosmological parameters is considered in the Planck scale.\\
Using \eqref{9} and \eqref{10} in \eqref{7} one can find the field equation as follows
\begin{equation}
\label{11}
3H^2=\frac{1}{2f_Q}\left(-\rho+\frac{f}{2}\right),
\end{equation}
\begin{equation}
\label{12}
\dot{H}+3H^2+\frac{\dot{f_Q}}{f_Q}H=\frac{1}{2f_Q}\left(\bar{p}+\frac{f}{2}\right),
\end{equation}
where dot $(^.)$ represents derivative with respect to $t$. The energy conservation equation for the viscous fluid can be written as
\begin{equation}
\label{13}
\dot{\rho}+3H(\rho+\bar{p})=0.
\end{equation}
Using equation \eqref{11} and \eqref{12}, we can find the following expressions
\begin{equation}
\label{14}
\rho=\frac{f}{2}-6H^2f_Q
\end{equation}
\begin{equation}
\label{15}
p=\left(\dot{H}+3H^2+\frac{\dot{f_Q}}{f_Q}H\right)2f_Q-\frac{f}{2}+3 \lambda H
\end{equation}
The equation of state (EoS) parameter can be written as
\begin{equation}
\label{16}
\omega=\frac{p}{\rho}.
\end{equation}

Now, one can use the above set up to explore the cosmological evolution  of the universe applying various approach.

Further, in analogy with GR, we can rewrite Eq.\eqref{11}, \eqref{12} as
\begin{equation}
3H^2=-\frac{1}{2}\rho_{eff}\,,
\end{equation}
\begin{equation}
\dot{H}+3H^2=\frac{p_{eff}}{2}\,.
\end{equation}
where
\begin{equation}
\label{ec1}
\rho_{eff}=\frac{1}{f_Q}\left(\rho-\frac{f}{2}\right),
\end{equation}
\begin{equation}
\label{ec2}
p_{eff}=-2\frac{\dot{f_Q}}{f_Q}H+\frac{1}{f_Q}\left(\bar{p}+\frac{f}{2}\right)
\end{equation}
Here, $p_{eff}$ and $\rho_{eff}$ are the effective pressure and energy density of the fluid content, respectively. The previous equations are going to be components of a modified energy-momentum tensor $T^{eff}_{\,\mu\nu}$, embedding the dependence on the trace of the nonmetricity tensor.

\section{Energy Conditions}\label{sec3}

The energy conditions (ECs) are the essential tools to understand the geodesics of the Universe. Such conditions can be derived from the well-known Raychaudhury equations, whose forms are \cite{Raychaudhuri/1955,10a,10b}
\begin{equation}
\label{17}
\frac{d\theta}{d\tau}=-\frac{1}{3}\theta^2-\sigma_{\mu\nu}\sigma^{\mu\nu}+\omega_{\mu\nu}\omega^{\mu\nu}-R_{\mu\nu}u^{\mu}u^{\nu}\,,
\end{equation}
\begin{equation}
\label{18}
\frac{d\theta}{d\tau}=-\frac{1}{2}\theta^2-\sigma_{\mu\nu}\sigma^{\mu\nu}+\omega_{\mu\nu}\omega^{\mu\nu}-R_{\mu\nu}n^{\mu}n^{\nu}\,,
\end{equation}
where $\theta$ is the expansion factor, $n^{\mu}$ is the null vector, and $\sigma^{\mu\nu}$ and $\omega_{\mu\nu}$ are, respectively, the shear and the rotation associated with the vector field $u^{\mu}$. In the Weyl geometry with the presence of non-metricity, the Raychaudhury equations takes different forms [see details calculations for the Raychaudhury equations with the non-metricity \cite{sa}]. For attractive gravity, equations \eqref{17}, and \eqref{18} satisfy the following conditions
\begin{align}
\label{19}
R_{\mu\nu}u^{\mu}u^{\nu}\geq0\,,\\
 R_{\mu\nu}n^{\mu}n^{\nu}\geq0\,.
\end{align}
 Therefore, if we are working with a perfect fluid matter distribution, the energy conditions for $f(Q)$ gravity are given by \cite{Mandal/2020},
\begin{itemize}
\item Strong energy conditions (SEC) if  $\rho_{eff}+3p_{eff}\geq 0\,$;
\item Weak energy conditions (WEC) if  $\rho_{eff}\geq 0, \rho_{eff}+ p_{eff}\geq 0\,$;
\item Null energy condition (NEC) if  $\rho_{eff}+3p_{eff}\geq 0\,$;
\item Dominant energy conditions (DEC) if $\rho_{eff}\geq 0, |p_{eff}|\leq \rho\,$.
\end{itemize}
Taking Eqs.  $\eqref{ec1}$ and $\eqref{ec2}$  into WEC, NEC and DEC constraints, we are able to prove that
\begin{itemize}
\item Weak energy conditions (WEC) if  $\rho\geq 0, \rho+p\geq 0\,$;
\item Null energy condition (NEC) if  $\rho+p\geq 0\,$;
\item Dominant energy conditions (DEC) if $\rho\geq 0, |p|\leq \rho\,$.
\end{itemize}
corroborating with the work from Capozziello et al.\cite{Capozziello/2018}. In the case of SEC condition, we yield to the constraint
\begin{equation}\label{20}
\rho+3\,p-9\lambda H-6\,\dot{f_Q}\,H+f \geq 0\,.
\end{equation}

Now, using above energy conditions, we can test the viability of our cosmological models. Further, it will helps us to understand our universe in  more realistic way.

\section{Data Interpretation}\label{sec4}

In this section, we adopted the parametrization technique, which will be used to reconstruct the cosmological models. For instance, one can see some interesting study, where they used parametrization technique to explore the cosmological models \cite{R1,R2}. The main advantage of adopting this technique is that we can study the cosmological models with observational data. As we know, the relation between the scale factor $a(t)$ and the redshift $z$ is given by $1+z=\frac{a_0}{a}$, where $a_0$ is the late time scale factor. From the above relation we can find $\frac{d}{dt}=-H(1+z)\frac{d}{dz}$. The non-metricity $Q$ in term of redshift $z$ can be written as
$Q=6 H_0^2 h(z)$, where $H(z)^2=H_0^2 h(z)$, and the late time Hubble parameter  $H_0=67.4\pm 0.5 km s^{-1}Mpc^{-1}$ \cite{Planck/2018}.

The interesting work of Sahni et al. \cite{Sahni/2003, e2} motivates us to take the functional form of $h(z)$ as follow;

\begin{equation}
\label{21}
h(z)=A_0+A_1(1+z)+A_2 (1+z)^2+A_3 (1+z)^3,
\end{equation}

where $A_0, \,\  A_1,\,\ A_2 $, and $A_3$ are constants. These constants can be measured by using the observational data. Also, we can find an additional constraint on the parameters for $z=0$ as $A_0+A_1+A_2+A_3=1$.

\subsection{Hubble Dataset}

Recently, a list of 57 data points of Hubble parameter in the redshift range $0.07\leq z\leq 2.41$ were compiled by Sharov and Vasiliev \cite{Sharov/2018}. This H(z) dataset was measured from the line-of-sight BAO data \cite{BAO1, BAO2, BAO3, BAO4, BAO5} and the differential ages $\Delta t$ of galaxies \cite{h1,h2,h3,h4}. The complete list of datasets is presented in \cite{Sharov/2018}. To estimate the model parameters, we used the Chi-square test by MCMC simulation. The Chi-square function is given by

\begin{equation}
\chi^2_{OHD}(p_s)=\sum_{i=1}^{57}\frac{[H_{th}(p_s,z_i)-H_{obs}]^2}{\sigma^2_{H(z_i)}},
\end{equation}

where $H_{obs}(z_i)$ represents the observed Hubble parameter values, $H_{th}(p_s,z_i)$ represents the Hubble parameter with the model parameters, and $\sigma^2_{H(z_i)}$ is the standard deviation.

\subsection{Pantheon Dataset}
Here, we use the 
latest Pantheon supernovae type Ia sample, which contains 1048 SNe Ia data points from SNLS, SDSS, Pan-STARRS1, HST surveys, and low-redshift in the redshift-range $z \in [0.01,2.3]$ to constraint the above parameters \cite{Scolnic/2018}. The $\chi^2_{SN}$ function from the Pantheon sample of 1048 SNe Ia \cite{Scolnic/2018} is given by
\begin{equation}
\chi^2_{SN}(p_1,....)=\sum_{i,j=1}^{1048}\bigtriangledown\mu_{i}\left(C^{-1}_{SN}\right)_{ij}\bigtriangledown\mu_{j},
\end{equation}
where $p_j$ represents the free parameters of the presumed model and $C_{SN}$ is the covariance metric \cite{Scolnic/2018}, and $\mu$ represents the distance moduli is given by;
 \begin{align*}
 \mu^{th}(z)& =5\log\frac{D_L(z)}{10pc},\, \quad D_L(z)=(1+z)D_M,\\
 D_M(z)&=c \int_0^{z}\frac{d\tilde{z}}{H(\tilde{z})},\, \quad \bigtriangledown\mu_{i}=\mu^{th}(z_i,p_1,...)-\mu_i^{obs}.
 \end{align*}
 
\begin{figure}[H]
\begin{center}
\includegraphics[scale=0.45]{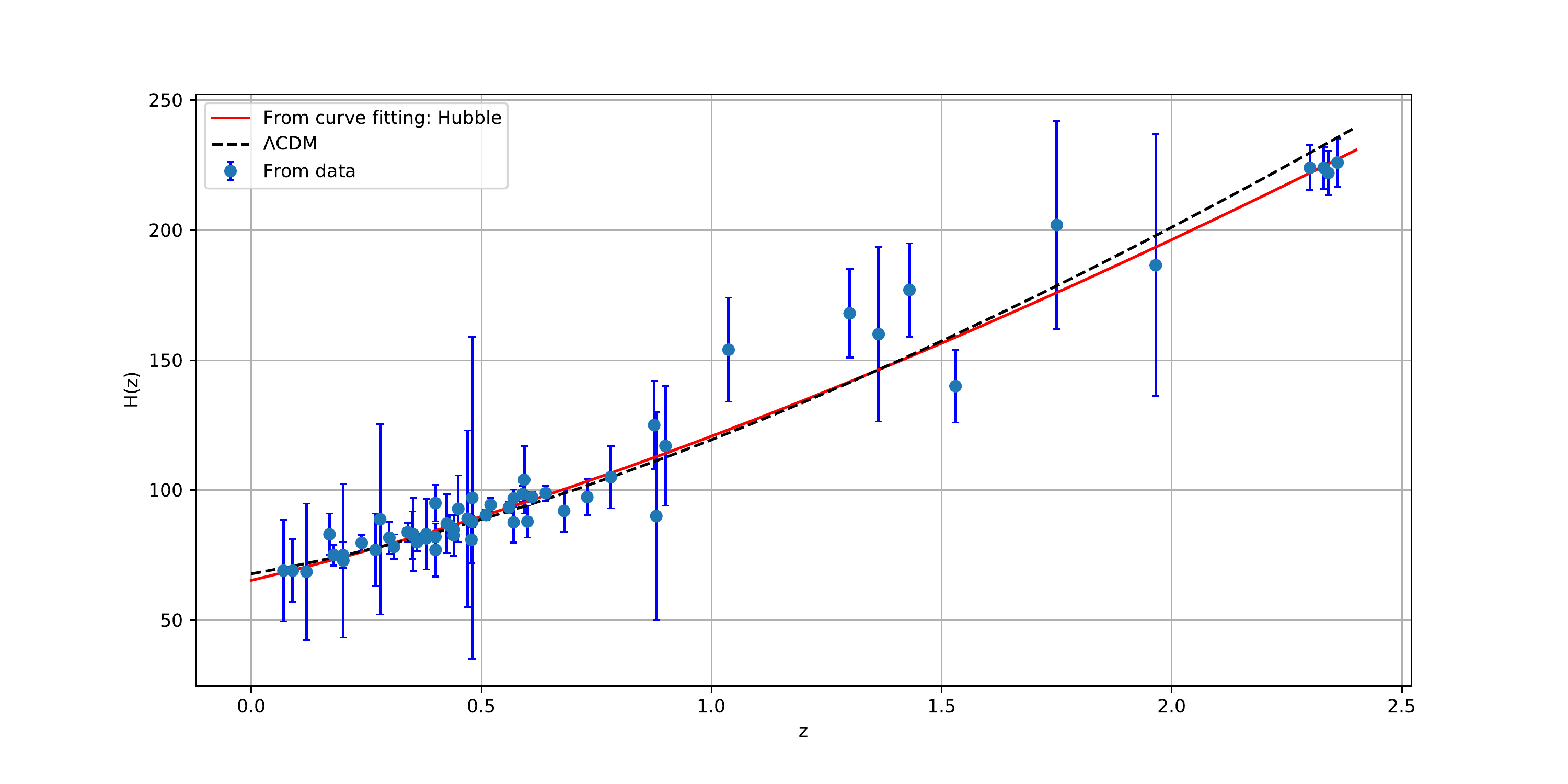}
\caption{The evolution of Hubble parameter $H(z)$ with respect to redshift $z$ is shown here. The red line represents our model (i.e., the profile of $H(z)$ as presented in \eqref{21}) and dashed-line indicates the $\Lambda$CMD model with $\Omega_{m0}=0.3$ and $\Omega_{\Lambda 0}=0.7$. The dots are shown the Hubble dataset with error bar. }
\label{f1a}
\end{center}
\end{figure}

\begin{figure}[H]
\begin{center}
\includegraphics[scale=0.45]{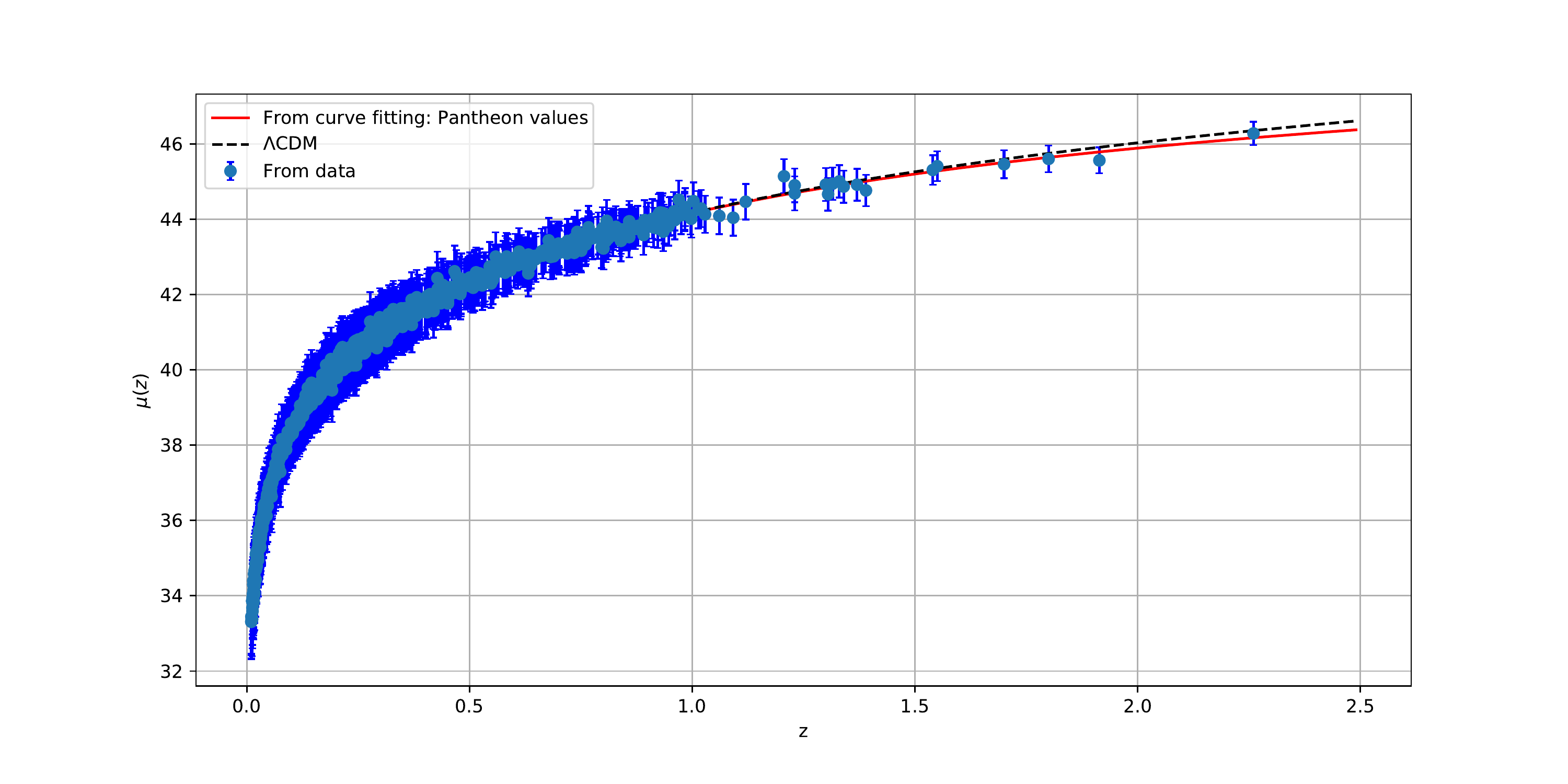}
\caption{The evolution of $\mu(z)$ with respect to redshift $z$ is shown here. The red line represents our model (i.e., the profile of $H(z)$ as presented in \eqref{21}) and dashed-line indicates the $\Lambda$CMD model with $\Omega_{m0}=0.3$ and $\Omega_{\Lambda 0}=0.7$. The dots are shown the 1048 Pantheon dataset with error bar. }
\label{f1c}
\end{center}
\end{figure}

\begin{figure}[H]
\includegraphics[scale=0.75]{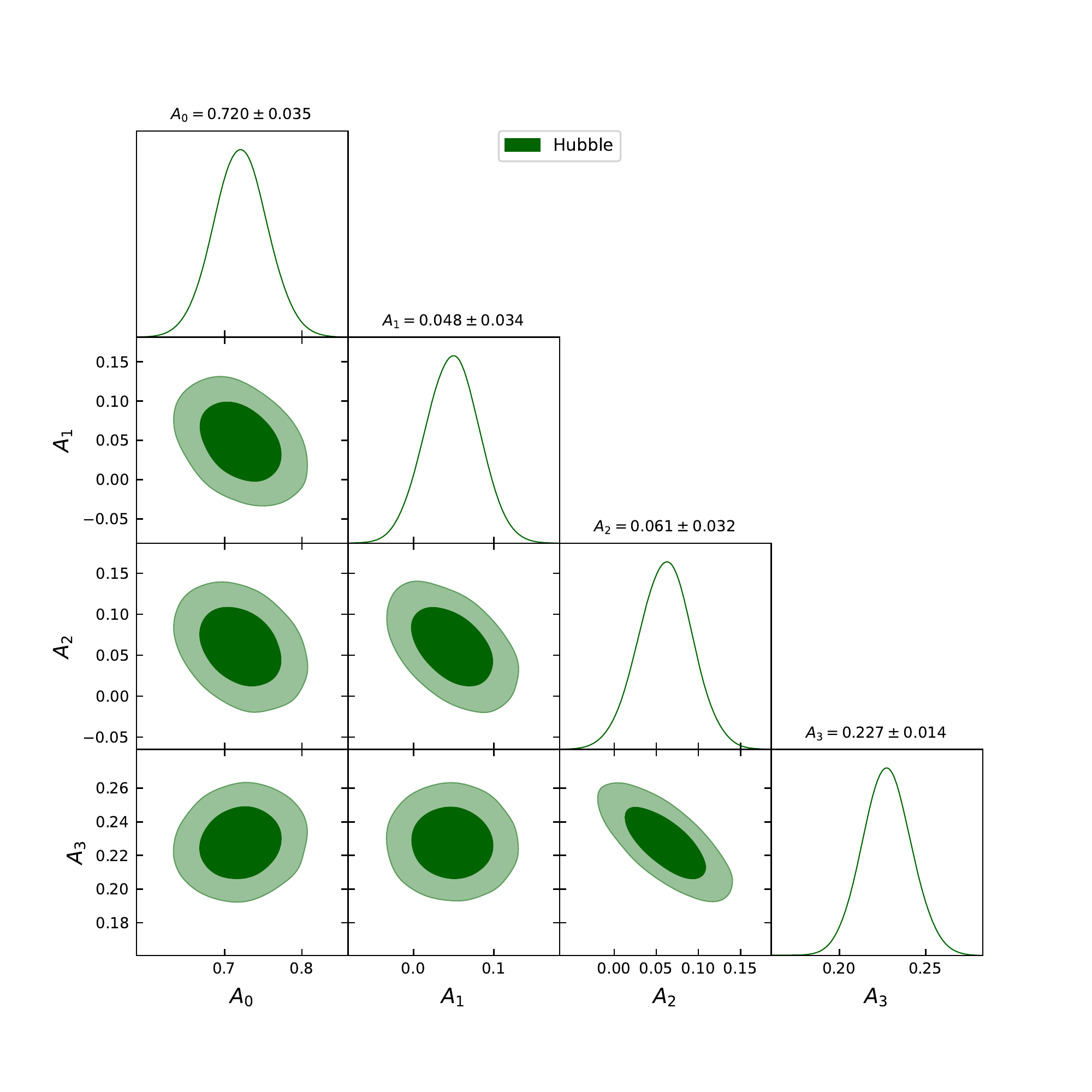}
\caption{The marginalized constraints on the coefficients in the expression of Hubble parameter $H(z)$ in Eqn. \eqref{21} are shown by using the Hubble sample}
\label{f1b}
\end{figure}

\begin{figure}[H]
\includegraphics[scale=0.75]{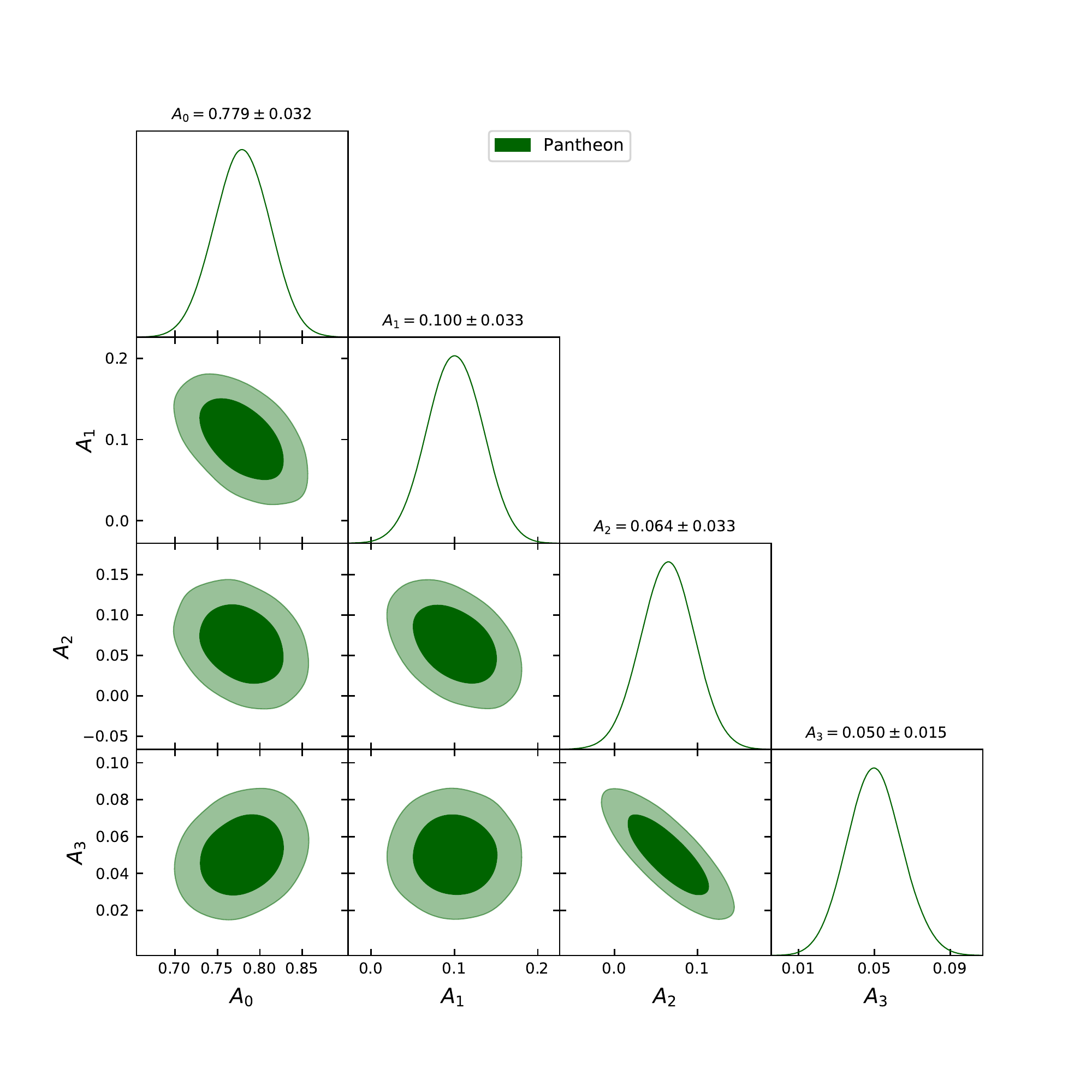}
\caption{The marginalized constraints on the coefficients in the expression of Hubble parameter $H(z)$ in Eqn. \eqref{21} are shown by using the Pantheon SNe Ia sample}
\label{f1d}
\end{figure}

We employ the emcee package in Python for performing a Markov chain Monte Carlo (MCMC) analysis, and provide the best-fit estimates and $2-\sigma $ upper limit of the parameters. We are presented the estimated values of parameters in Table \ref{t1}.
\begin{table}[H]
	\renewcommand\arraystretch{1.5}
	\caption{The marginalized constraining results on four model parameters are shown by using the Hubble and Pantheon SNe Ia sample.
	}
	\begin{center}
	\begin{tabular} { l |c| c  }
		\hline
		\hline

		Dataset    &H(z) dataset &Pantheon dataset \\
 \hline
 $A_0$  & $0.720\pm 0.035$  & $0.779\pm 0.032$\\
 \hline
 $A_1$  & $0.048\pm 0.034$ & $0.100\pm 0.033$  \\
 \hline
  $A_2$ & $0.061\pm 0.032$  & $0.064\pm 0.033$  \\
 \hline
 $A_3$  & $0.227\pm 0.014$  & $0.050\pm 0.015$\\

	    \hline
		\hline
	\end{tabular}
	\end{center}
	\label{t1}

\end{table}
Refer to the triangle plot in Fig. \ref{f1b} and \ref{f1d} for a complete survey of the parameter space with respect to Hubble and Pantheon data sample; the values are restricted to the positive quadrant as they behave like the density parameters. For further study, we refer to these constrained values of the parameters. Furthermore, in Figure \ref{f1a} and \ref{f1c}, we present the datasets such as Hubble and Pantheon sample with our model, respectively.

\section{Viscous Fluid Models in $f(Q)$ gravity}\label{sec5}

Here, we are going to discuss the cosmological models constructed using the bulk viscous fluid. We also test the energy conditions of viscous fluid models to check their self-stability. We have constructed three models; in the first case, we presume a linear functional form of $Q$. The motivation behind taking this form is that it recovers the fundamental laws of gravity. Besides this, we discuss the power-law form of $Q$ and logarithmic dependence of $f(Q)$ models in the second and third case, respectively.

\subsection{Model-1: $f(Q)=\alpha Q$}
In this subsection, we assume the linear functional form of $Q$ with the free parameter $\alpha<0$. Now, using $f(Q)=\alpha Q$, we find the following expressions for energy density, pressure, and equation of state parameter, respectively.

\begin{equation}\label{22}
\rho=-3 \alpha H_0^2 h(z)
\end{equation}
\begin{equation}\label{23}
p=\frac{H_0 \left(\alpha  h'(z)+3 \alpha  H_0 h(z)^{3/2}+3 \lambda  h(z)\right)}{\sqrt{h(z)}}
\end{equation}

\begin{equation}\label{24}
\omega=-1-\frac{\alpha  h'(z)}{3 \alpha  H_0h(z)^{3/2}}-\frac{3 \lambda }{3 \alpha  H_0\sqrt{h(z)}}
\end{equation}

\begin{figure}[H]
\begin{center}
\includegraphics[width=6 cm]{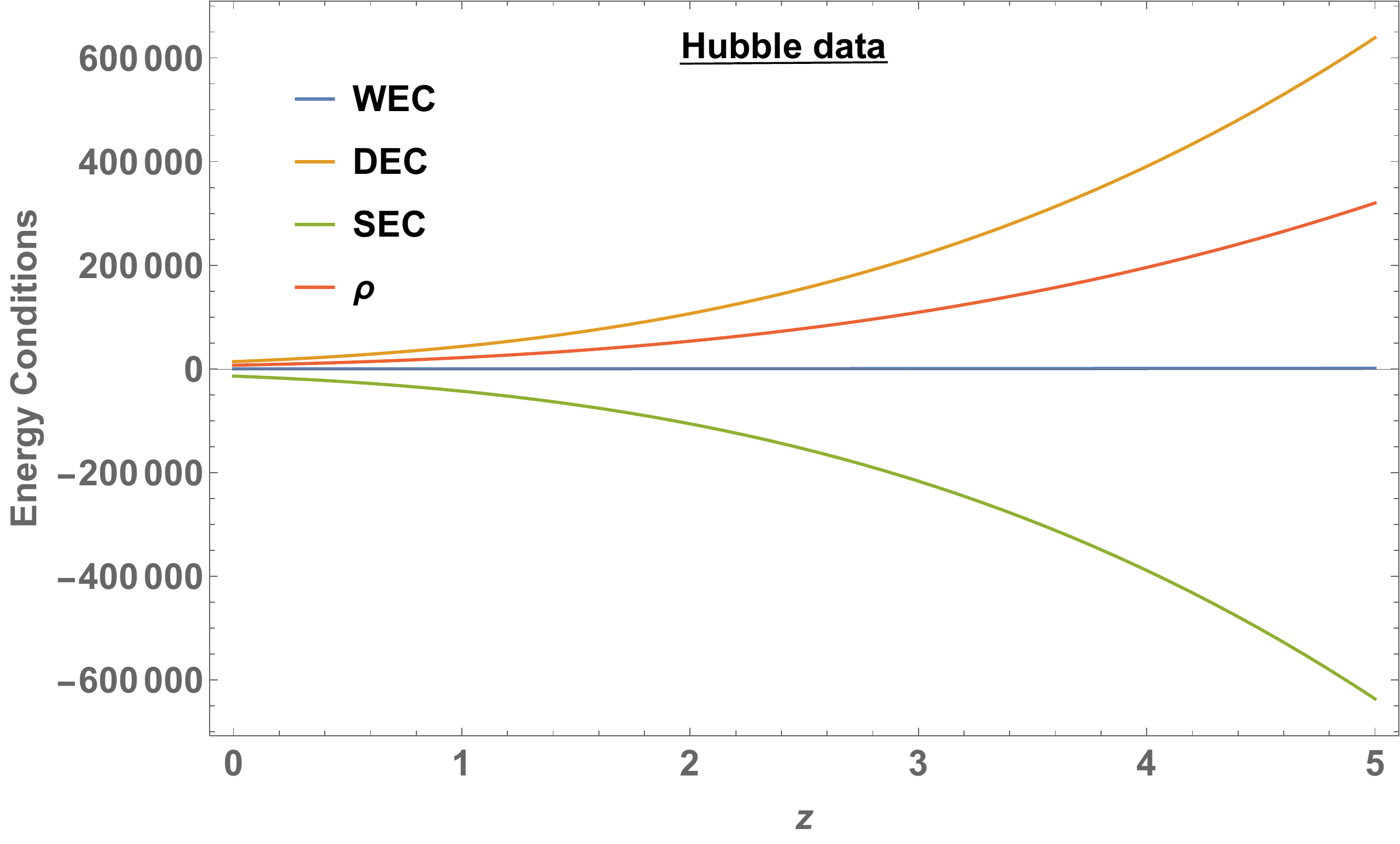}\hspace*{0.5 cm}
\includegraphics[width=6 cm]{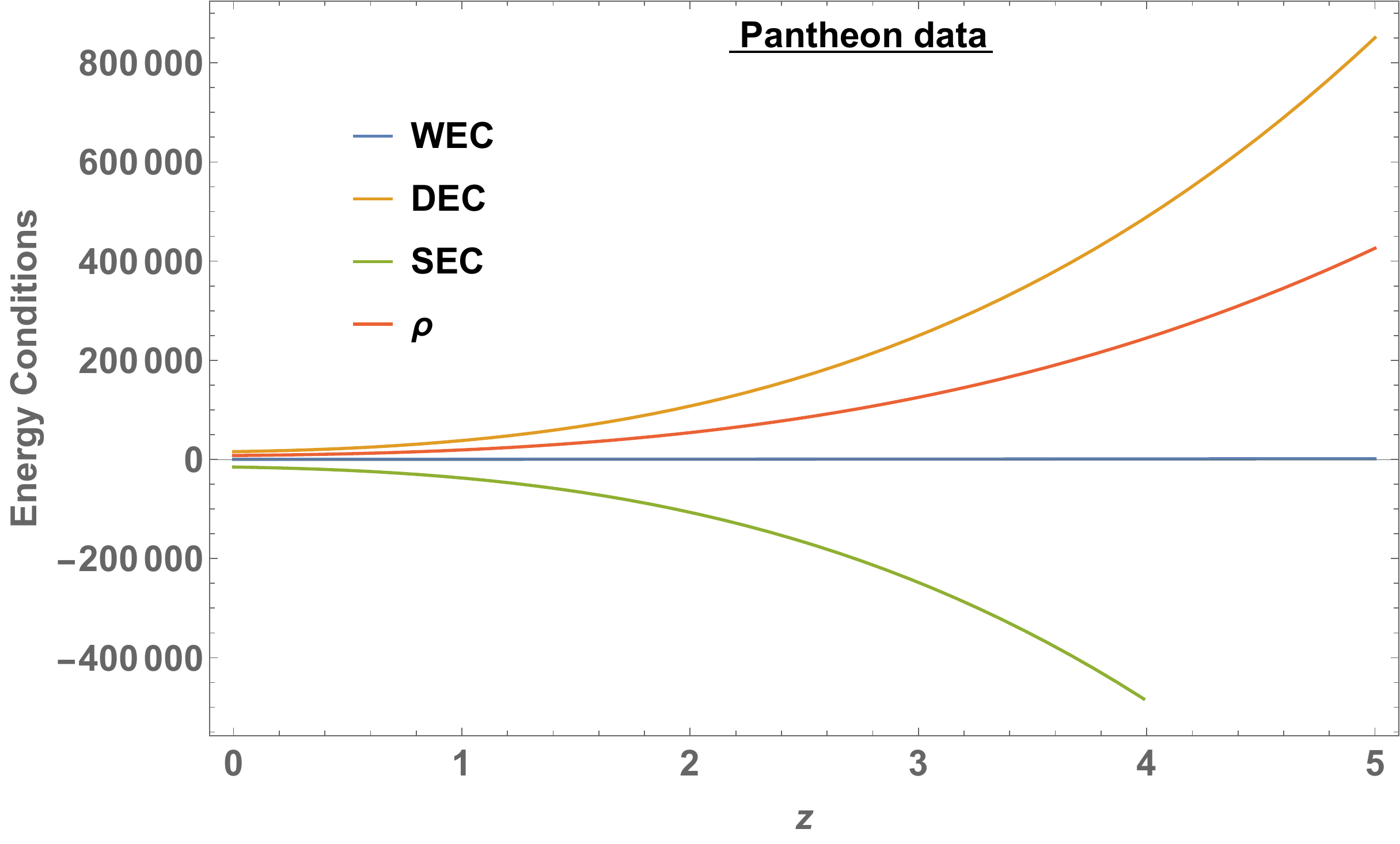}
\caption{The behavior of energy conditions against the redshift parameter $z$ with the constraint values of the coefficients in  \eqref{22} $\lambda=1$ and $\alpha=-0.5$ for $f(Q)=\alpha Q$.}
\label{f2}

\end{center}
\end{figure}

\begin{figure}[H]
\begin{center}
\includegraphics[width=6 cm]{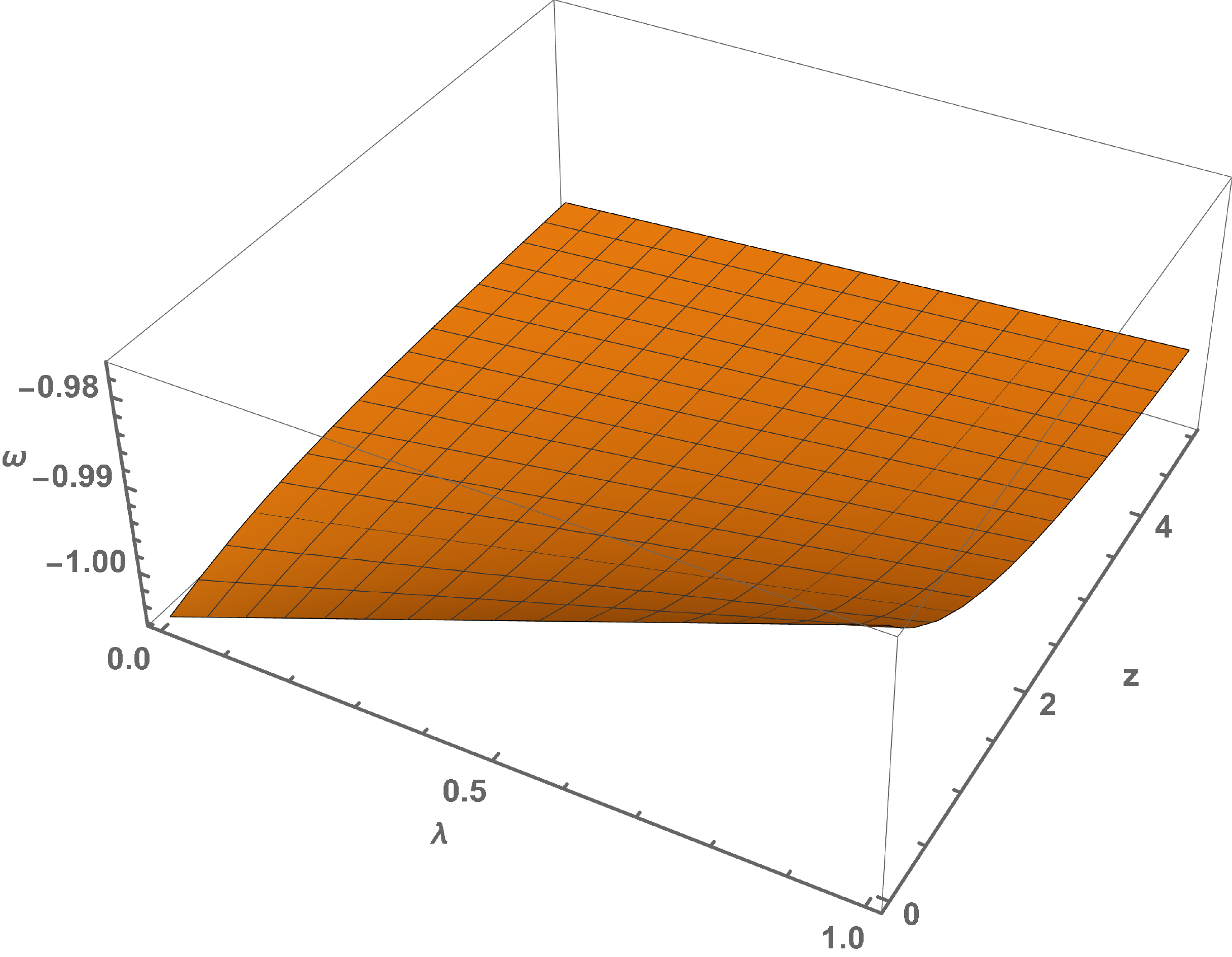}\hspace*{0.5 cm}
\includegraphics[width=6 cm]{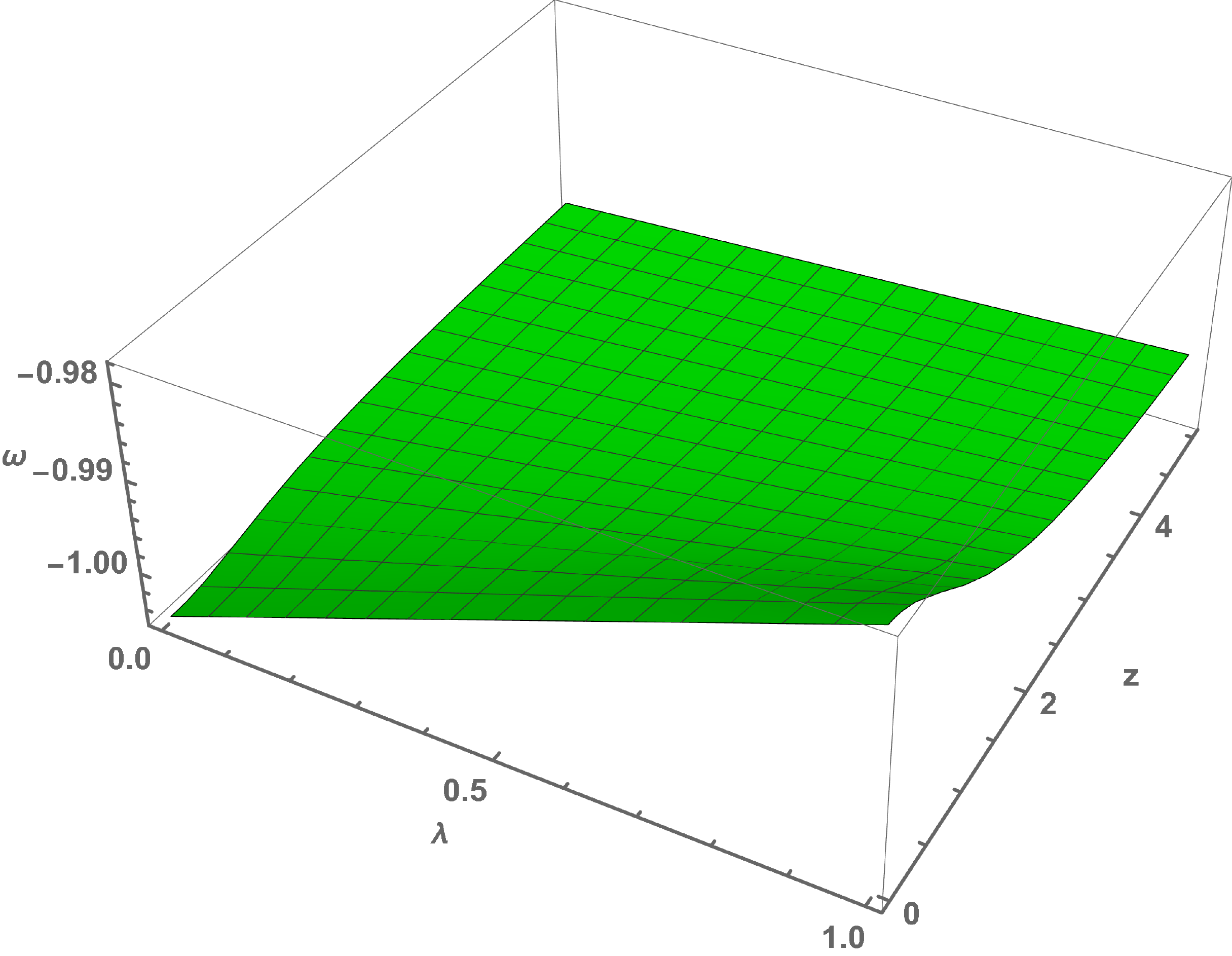}
\caption{The behavior of EoS parameter against the redshift parameter $z$ and $\lambda$ with the constraint values of the coefficients in  \eqref{22} and $\alpha=-0.5$ for $f(Q)=\alpha Q$. \textit{Left side }graph for Hubble data and \textit{right side }graph for Pantheon data.}
\label{f3}
\end{center}
\end{figure}

Using \eqref{23} and \eqref{24} in the energy conditions, we have shown the profiles of energy density, WEC, NEC, DEC, and SEC against redshift $z$ in Fig. \ref{f2}. From that figure, one can clearly observe that all the energy conditions satisfy while the SEC is violated. These are in agreement with the present scenario of the universe. In Fig. \ref{f3}, we have drawn the equation of state parameter behavior concerning the redshift $z$ and $\lambda$. The profile of EoS shows that it takes its values very close to $-1$, which aligns with the result of the $\Lambda$CDM model. Further, we observed that for negative values of $\lambda$, our model is showing the phantom behaviour.

\subsection{Model-2: $f(Q)= Q+ m Q^n$}
\justifying
For second model, we presume a power-law functional form of $Q$, where $m$ and $n$ are the free model parameters \cite{Mandal/2020}. For this, the energy density, pressure, and equation of state parameter can be rewritten as

\begin{figure}[H]
\begin{center}
\includegraphics[width=6 cm]{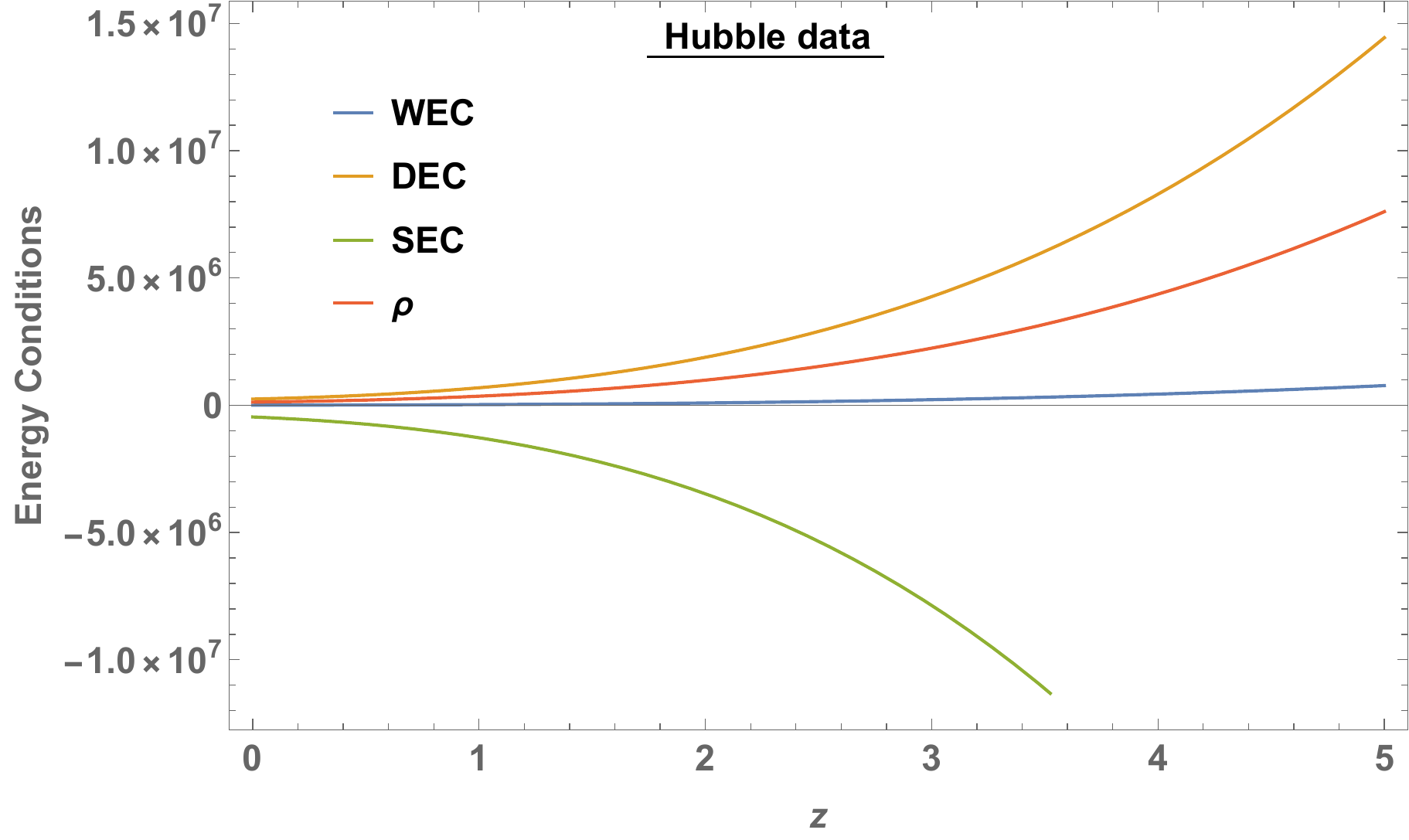}\hspace*{0.5 cm}
\includegraphics[width=6 cm]{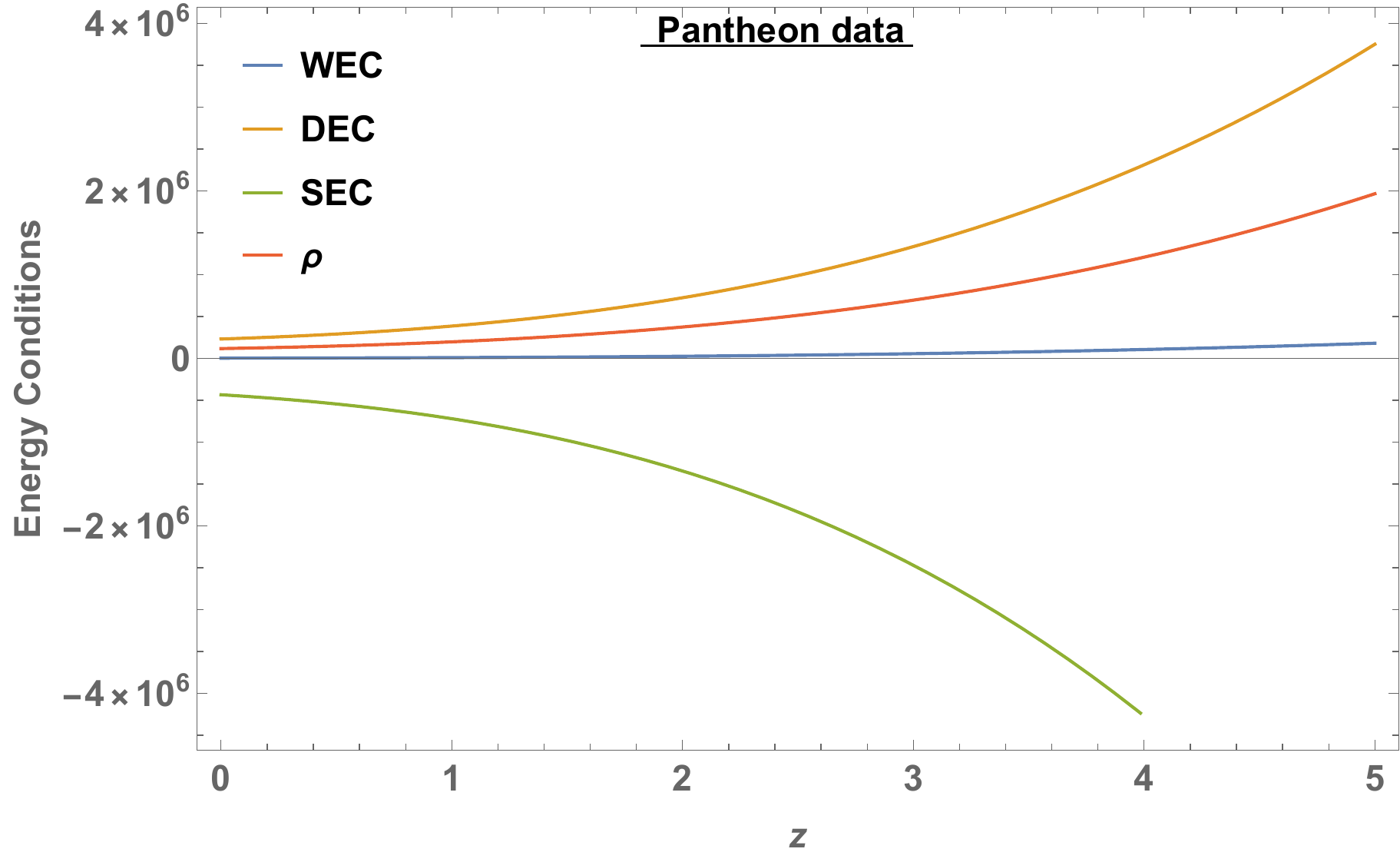}
\caption{The behavior of energy conditions against the redshift parameter $z$ with the constraint values of the coefficients in  \eqref{22} $\lambda=2 ,\,\ m=-5 ,\,\ n=1.05$ for $f(Q)= Q+m Q^n$}
\label{f4}
\end{center}
\end{figure}

\begin{equation}\label{25}
\rho=\frac{1}{2} \left(m \left(-6^n\right) (2 n-1) \left(H_0^2 h(z)\right)^n-6 H_0^2 h(z)\right)
\end{equation}

\begin{multline}\label{26}
p=\frac{m 6^{n-1} n h'(z) \left(H_0^2 h(z)\right)^n}{H_0 h(z)^{3/2}}-\frac{m 2^n 3^{n-1} (n-1) n (z+1) h'(z) \left(H_0^2 h(z)\right)^n}{h(z)}+\frac{H_0 h'(z)}{\sqrt{h(z)}}\\
+m 2^{n-1} 3^n (2 n-1) \left(H_0^2 h(z)\right)^n+3 H_0^2 h(z)+3 H_0 \lambda  \sqrt{h(z)}
\end{multline}

\begin{multline}\label{27}
\omega=\frac{-m 6^n n h'(z) \left(H_0^2 h(z)\right)^n+H_0 m 2^{n+1} 3^n (n-1) n (z+1) \sqrt{h(z)} h'(z) \left(H_0^2 h(z)\right)^n-6 H_0^2 h(z) h'(z)}{3 h(z)^{3/2} \left(6 H_0^3 h(z)+H_0 m 6^n (2 n-1) \left(H_0^2 h(z)\right)^n\right)}\\
+\frac{-18 H_0^3 h(z)^{5/2}+H_0 m \left(-2^n\right) 3^{n+1} (2 n-1) h(z)^{3/2} \left(H_0^2 h(z)\right)^n-18 H_0^2 \lambda  h(z)^2}{3 h(z)^{3/2} \left(6 H_0^3 h(z)+H_0 m 6^n (2 n-1) \left(H_0^2 h(z)\right)^n\right)}
\end{multline}

\begin{figure}[H]
\begin{center}
\includegraphics[width=6 cm]{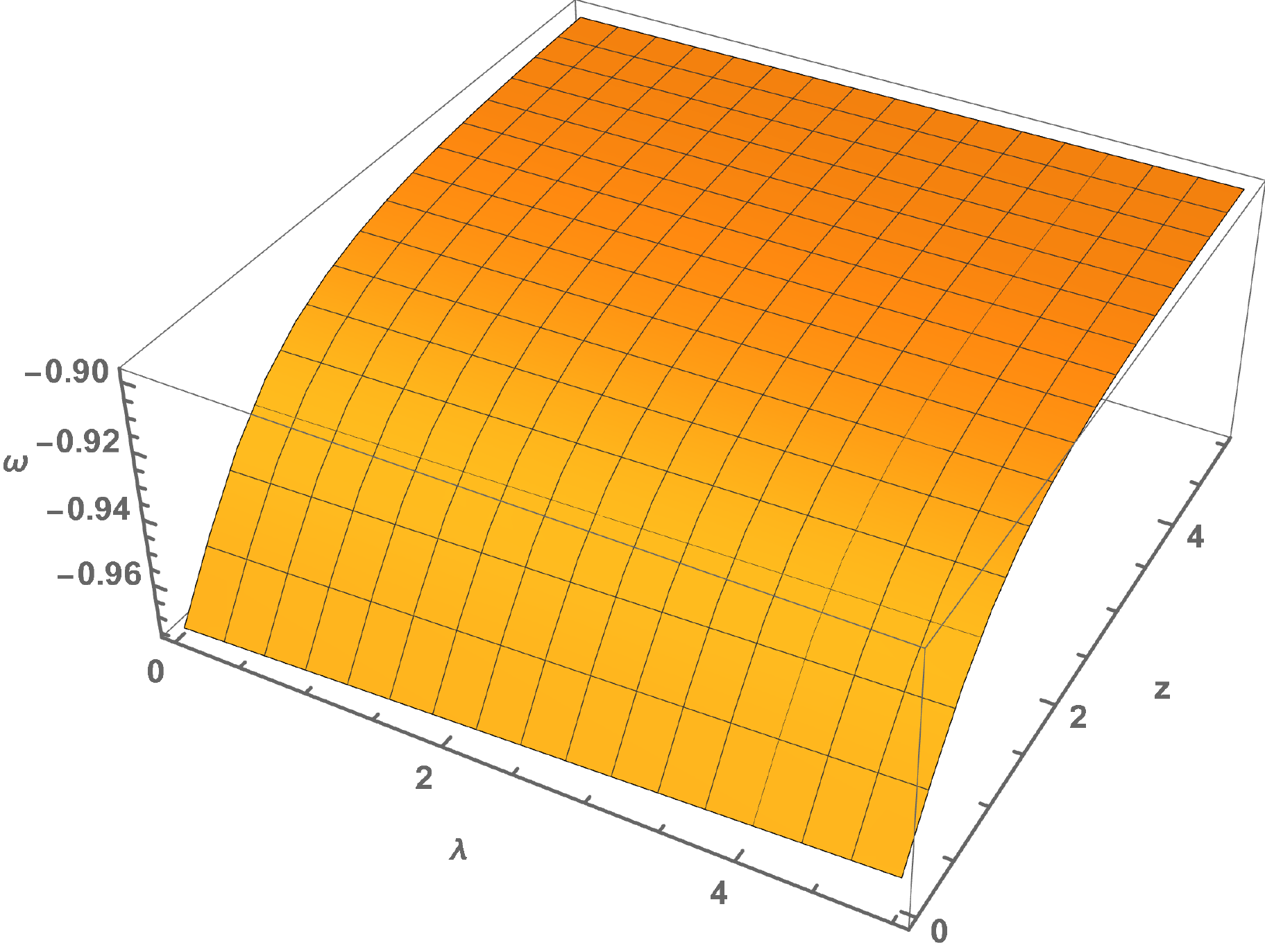}\hspace*{0.05 cm}
\includegraphics[width=6 cm]{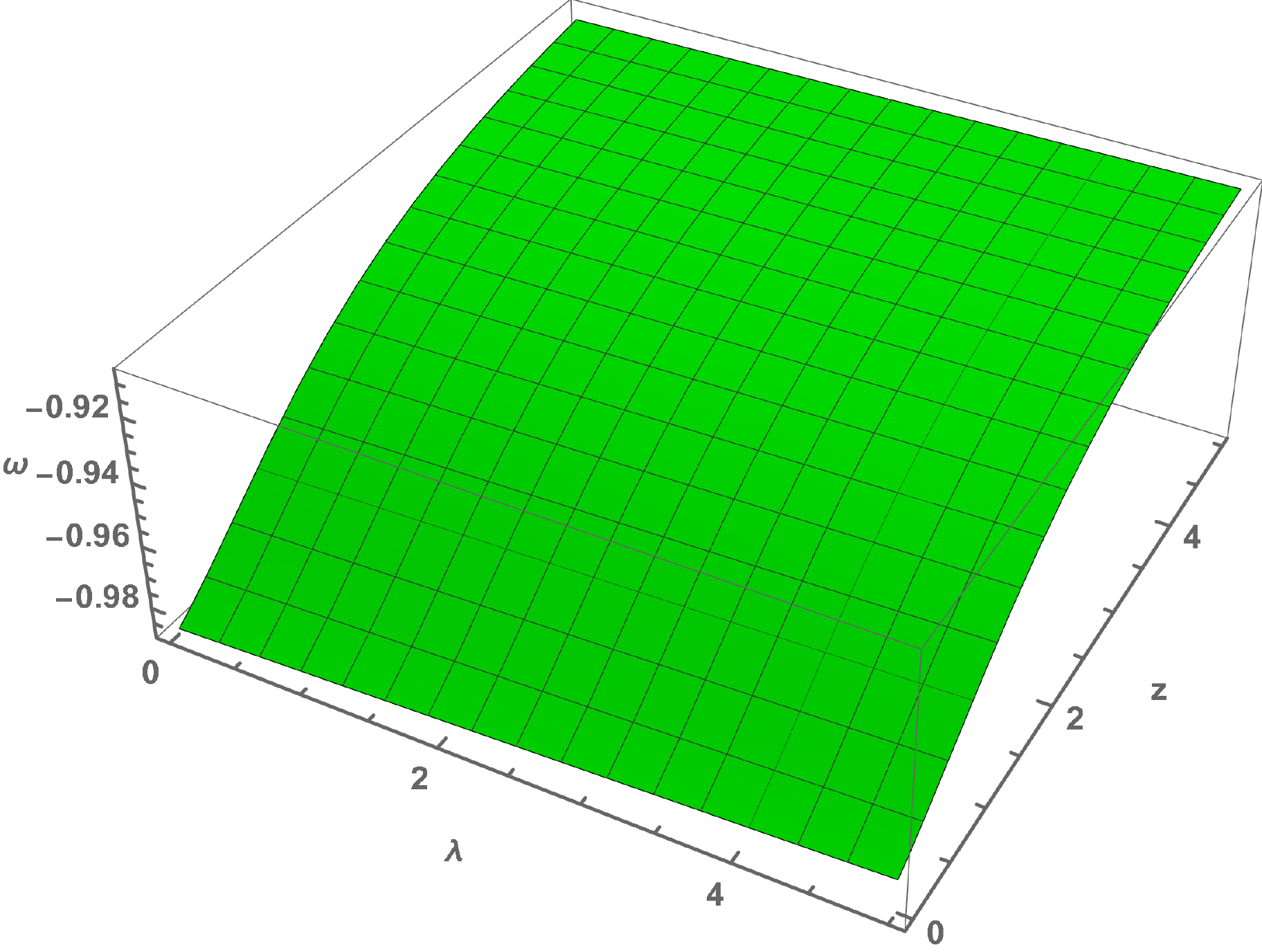}
\caption{The behavior of EoS parameter against the redshift parameter $z$ and $\lambda$ with the constraint values of the coefficients in  \eqref{22} and $ m=-5 ,\,\ n=1.05$   for $f(Q)= Q+m Q^n$.\textit{Left side }graph for Hubble data and \textit{right side }graph for Pantheon data.}
\label{f5}
\end{center}
\end{figure}

It is well-known that, the energy density should be non-negative ti have a viable cosmological model. So, keeping this in mind and using the constraint values of parameters, we found the relation between $m$ and $n$ as $m \,28566^n e^{0.0544882 n} (0.5\, -1. n)-15082.8\geq 0$ (for Hubble) and $m\, 28566^n e^{-0.00702461 n} (0.5\, -1. n)-14183\geq 0$ (for Pantheon) such that $\rho\geq 0$. In Fig. \ref{f4}, the profiles of all the energy conditions have been drawn using \eqref{26} and \eqref{27} in the above conditions. From those profiles, we conclude that all those behavior indicates the accelerated expansion of the universe, i.e., SEC has violated, and other energy conditions satisfied. The behavior EoS have been shown using \eqref{28} in Fig. \ref{f5}. Also, its values lie near to $-1$.

\subsection{model-3: $f(Q)=\gamma +\beta \log Q$}

Here, we discuss the logarithmic function of the non-metricity $Q$ having the free parameters $\gamma$ and $\beta$ \cite{Mandal/2020}. The energy density, pressure, and equation of state parameter can be written as 

\begin{equation}\label{28}
\rho=\frac{1}{2} \left(-2 \beta +\gamma +\beta  \log \left(6 H_0^2 h(z)\right)\right)
\end{equation}
\begin{equation}\label{29}
p=\frac{1}{6} \left(\frac{\beta  h'(z)}{H_0 h(z)^{3/2}}+\frac{2 \beta  (z+1) h'(z)}{h(z)}-3 \left(-2 \beta +\gamma +\beta  \log \left(6 H_0^2 h(z)\right)\right)+18 H_0 \lambda  \sqrt{h(z)}\right)
\end{equation}
\begin{equation}\label{30}
\omega=\frac{2 \beta  H_0 (z+1) \sqrt{h(z)} h'(z)+\beta  h'(z)-3 H_0 h(z)^{3/2} \left(-2 \beta +\gamma +\beta  \log \left(6 H_0^2 h(z)\right)\right)+18 H_0^2 \lambda  h(z)^2}{3 H_0 h(z)^{3/2} \left(-2 \beta +\gamma +\beta  \log \left(6 H_0^2 h(z)\right)\right)}
\end{equation}

\begin{figure}[H]
\begin{center}
\includegraphics[width=6 cm]{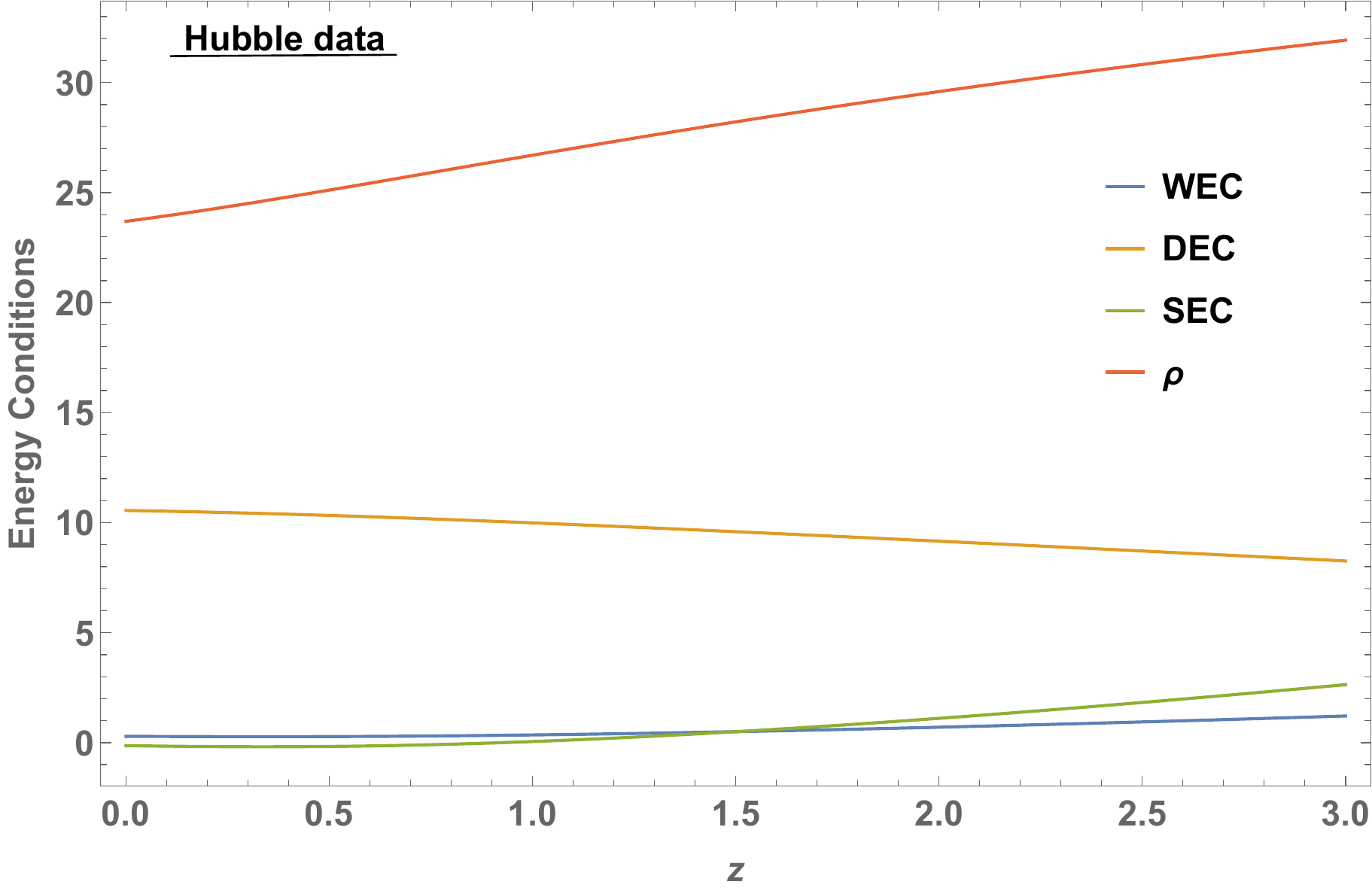}\hspace*{0.05 cm}
\includegraphics[width=6 cm]{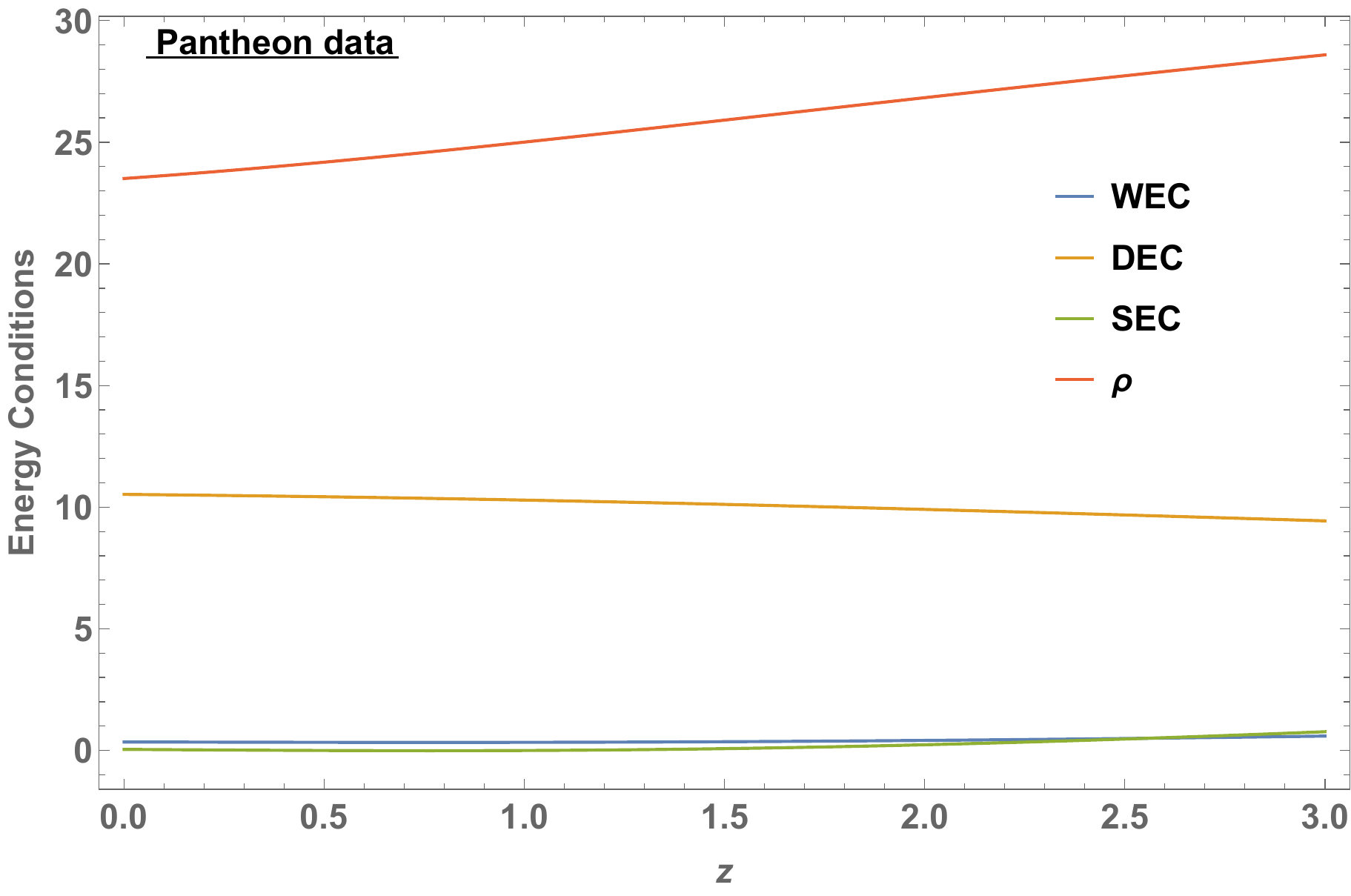}
\caption{The behavior of energy conditions against the redshift parameter $z$ with the constraint values of the coefficients in  \eqref{22}, for $\lambda=0.002 ,\,\ \gamma=15 ,\,\ \beta=-0.5$, and $f(Q)=\gamma+\beta \log Q$.}
\label{f6}
\end{center}
\end{figure}

\begin{figure}[H]
\begin{center}
\includegraphics[width=6 cm]{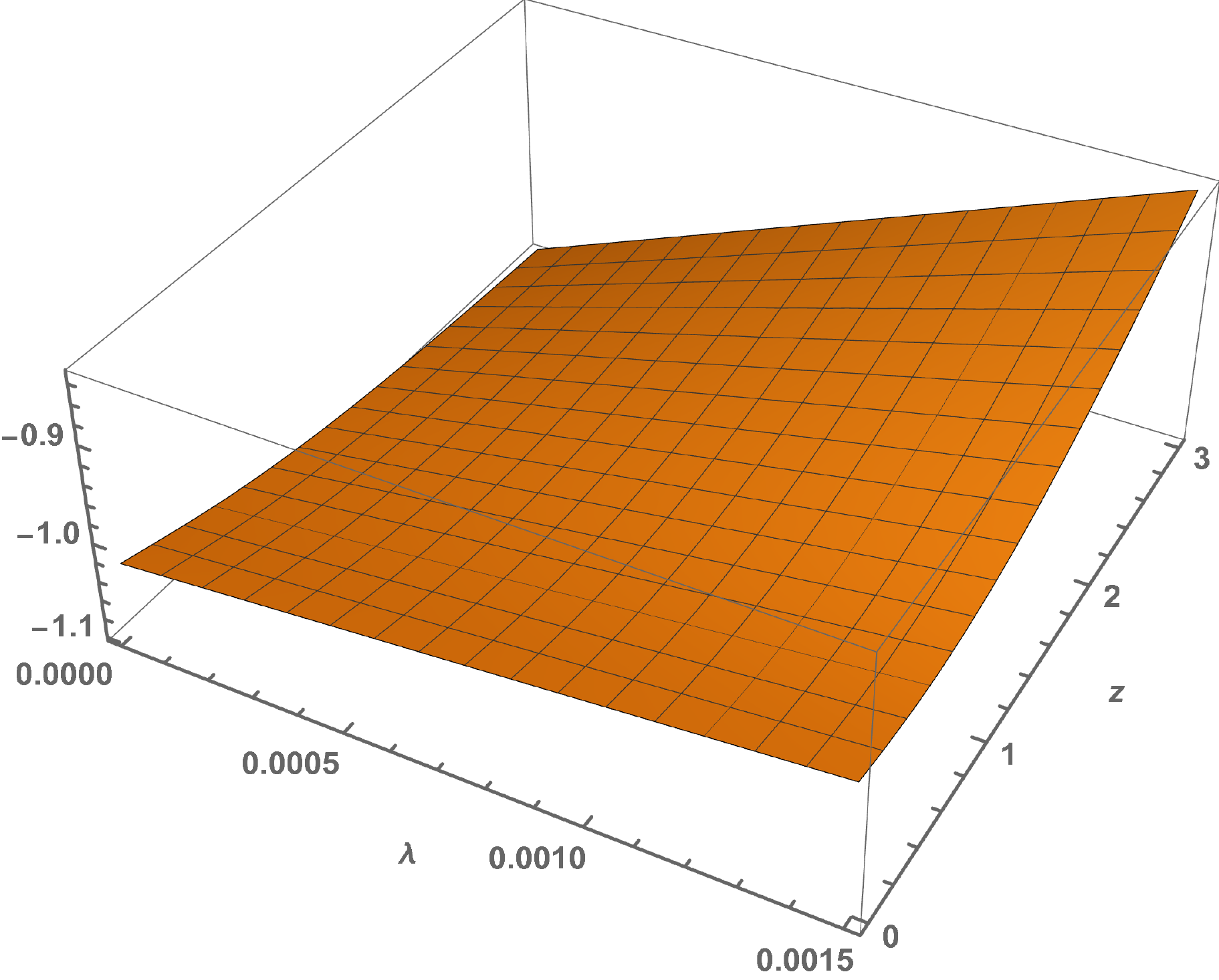}\hspace*{0.05 cm}
\includegraphics[width=6 cm]{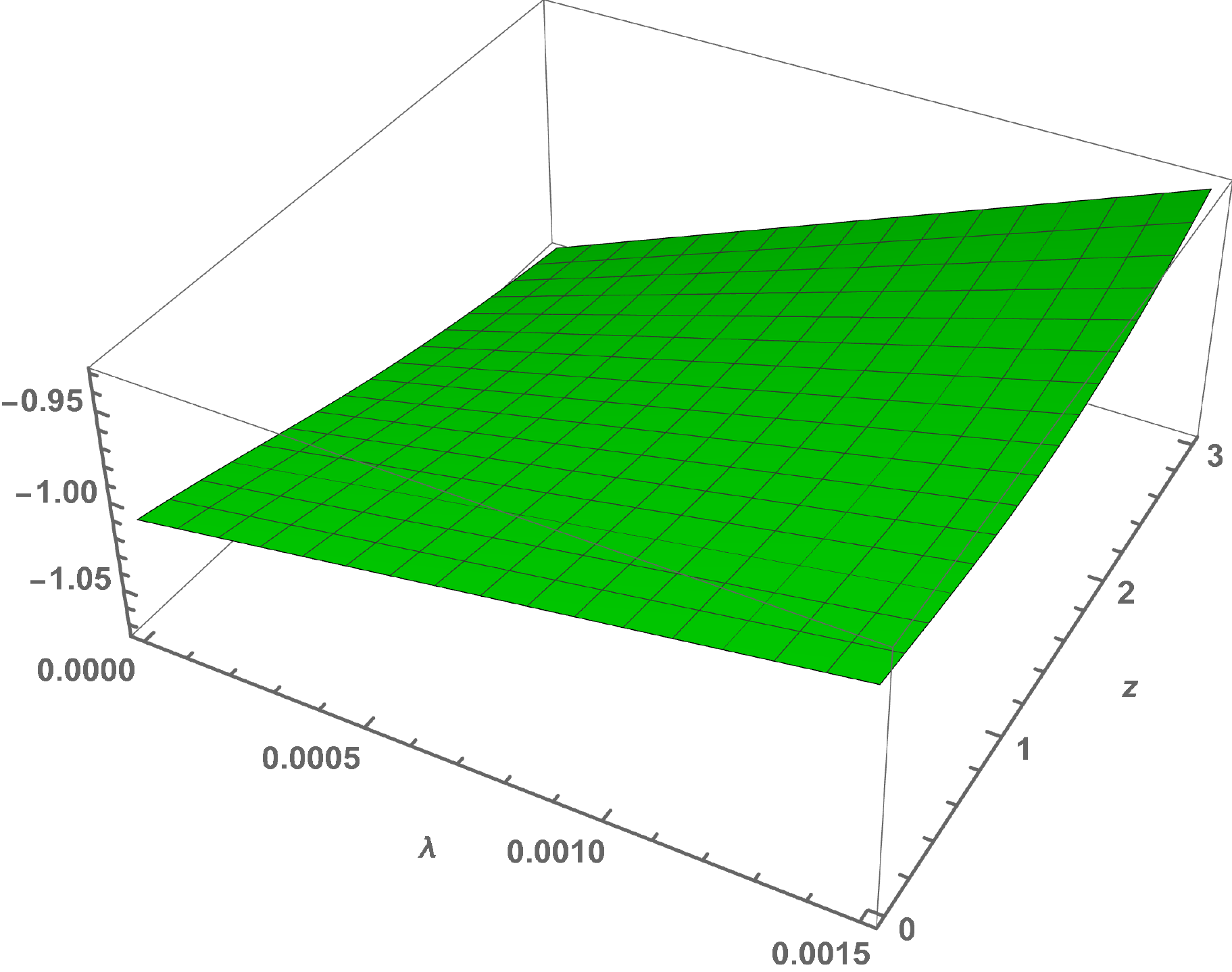}
\caption{The behavior of EoS parameter against the redshift parameter $z$ and $\lambda$ with the constraint values of the coefficients in  \eqref{22} and $\gamma=15 ,\,\ \beta=-0.5$ for $f(Q)=\gamma+\beta \log Q$. \textit{Left side }graph for Hubble data and \textit{right side }graph for Pantheon data.}
\label{f7}
\end{center}
\end{figure}

Following same procedure as discussed above $\rho\geq 0$, we found that $\gamma\geq -8.13 \beta$ (for Hubble) and $\gamma\geq -8.25 \beta $ (for Pantheon). In Fig. \ref{f6}, \ref{f7}, the profiles of energy conditions and equation of state parameter have been presented, respectively. For this case, SEC has been violated, and the energy density is positive over the whole range. This result is also an agreement with the accelerated expansion of the universe. The values of $\omega$ take close to $-1$.  

\section{Final Remarks}\label{sec6}

The recent interests in the universe impel us to go beyond the standard formulation of gravitational interaction, and for this, several modified theories of gravity have been proposed in the literature. However, one of the crucial roles is to define their self-stability. As it is well known that energy conditions are the best tool to test the cosmological models' self-consistency. The physical motivation to check the energy conditions of a new cosmological model helps us describe its compatibility with the space-time casual and geodesic structure. In this manuscript, we presumed a well motivated Hubble parameter and then constructed the cosmological models by adding bulk viscous fluid in the cosmological fluid within the $f(Q)$ gravity framework. We have also adopted the parametrization technique to discuss the null, the strong, the weak, and the dominated energy conditions for three types of $f(Q)$ gravity models. The Hubble dataset and largest Pantheon supernovae dataset were used to constraint the coefficients in the expression for the Hubble parameter.

In our first approach, we have considered a linear function of the non-metricity $Q$ model ($f(Q)\propto Q$). Such a model helps us to deal with the fundamental theories. The profiles are presented in Fig. \ref{f2} reveals the accelerated expansion of the universe. For our second model, we have considered a polynomial function of $Q$ having two free parameters m, and n. The self-stability of this model was checked through the energy conditions. From Fig. \ref{f4}, we observed that SEC violated while other energy conditions are satisfied. For the last model, we have presumed a logarithmic functional form of the non-metricity $Q$ with two free parameters $\gamma$, and $\beta$. The graphics depicted in Fig. \ref{f5} indicate the universe's accelerated expansion phase with a specific range of $\gamma$ and $\beta$. Moreover, such a model violates SEC with the positive energy density. This type of results are the good agreement for the current accelerated scenario of the universe.

For the shake of completeness, we derived and discussed the behaviors of equation of state parameter (EoS) for three viscous fluid models. In Fig. \ref{f3}, \ref{f5}, and \ref{f7}, the profiles of $\omega$ have been shown for three cases respectively. From those figures, one can observe that $\omega$ presenting its values very close to -1, which is compatible with the negative pressure of the present scenario of the universe. These results also collaborate with current astronomical observations, as well as the $\Lambda$CDM description for dark energy \cite{Planck/2018}. In addition, we observed that, for all models, the equation of state parameter $\omega$ converges to phantom  phase for negative values of viscous fluid parameter $\lambda$.

Further, one can compare the bulk viscosity effect on our three models. In the case of Model-1 and Model-2, any value of bulk viscosity coefficient $\lambda>0$ shows a consistent result with the observation and the current scenario of the universe. Whereas in the case of Model-3, $\lambda$ becomes very sensitive. Because if we will consider the value of $\lambda>10^{-2}$, then our DEC will violate. As a result, observed particles move faster than light, which leads to singularity in the present stage. In conclusion, we can say that the viscosity effect is more in the case of model-1 and model-2 in comparison to model-3.

The above results allowed us to examine the self-stability of the different families of bulk viscous fluid models in symmetric teleparallel gravity. Also, it sheds light on a new direction of modified theories compatible with the recent interests, particularly, the accelerated expansion of the universe. Moreover, it would be interesting to explore the symmetric teleparallel gravity in more generalized viscous fluid models. That may provide us some impressive results. In the near future, we plan to investigate some of the above ideas and hope to report them.

\section*{Acknowledgements}

S.M. acknowledges Department of Science \& Technology (DST), Govt. of India, New Delhi, for awarding INSPIRE Fellowship (File No. DST/INSPIRE Fellowship/2018/IF180676). PKS acknowledges CSIR, New  Delhi, India for financial support to carry out the Research project [No.03(1454)/19/EMR-II, Dt. 02/08/2019].

\section*{Data Availability Statement}

There are no new data associated with this article.

\end{document}